\documentclass[prl,showpacs,twocolumn,superscriptaddress]{revtex4-2}
\usepackage{graphicx}
\usepackage{amssymb}
\usepackage{amsfonts}
\usepackage{color}

\begin{document}

\title{Breakdown of Linear Spin-wave Theory and Existence of Spinon Bound States in the Frustrated Kagome Lattice Antiferromagnet}
\author{K.~Matan}
\email[]{kittiwit.mat@mahidol.ac.th}
\affiliation{Department~of~Physics,~Faculty~of~Science,~Mahidol~University, Bangkok 10400, Thailand}
\affiliation{ThEP,~Commission of Higher Education,~Bangkok 10400, Thailand}
\author{T.~Ono}
\affiliation{Department of Physical Science, School of Science, Osaka Prefecture University, Sakai, Osaka 599-8531, Japan}
\author{S. Ohira-Kawamura}
\affiliation{Materials and Life Science Division, J-PARC Center, Tokai, Ibaraki 319-1195, Japan}
\author{K.~Nakajima}
\affiliation{Materials and Life Science Division, J-PARC Center, Tokai, Ibaraki 319-1195, Japan}
\author{Y.~Nambu}
\affiliation{Institute for Materials Research, Tohoku University, Sendai 980-8577, Japan}
\affiliation{FOREST, Japan Science and Technology Agency, Kawaguchi, Saitama 332-0012, Japan}
\affiliation{Organization for Advanced Studies, Tohoku University, Sendai 980-8577, Japan}
\author{T.~J.~Sato}
\affiliation{Institute of Multidisciplinary Research for Advanced Materials, Tohoku University, 2-1-1 Katahira, Sendai, Miyagi 980-8577, Japan}

\date{\today}
\begin{abstract}

Spin dynamics of the spin-1/2 kagome lattice antiferromagnet Cs$_2$Cu$_3$SnF$_{12}$ was studied using high-resolution, time-of-flight inelastic neutron scattering.  The flat mode, a characteristic of the frustrated kagome antiferromagnet, and the low-energy dispersive mode, which is dominated by magnons, can be well described by the linear spin-wave theory.  However, the theory fails to describe three weakly dispersive modes between 9 and 14~meV.  These modes could be attributed to two-spinon bound states, which decay into free spinons away from the zone center and at a high temperature, giving rise to continuum scattering.
\end{abstract}
\maketitle

At zero temperature, magnetic moments in a strongly correlated electron system either order~\cite{Neel} or remain disordered but are highly entangled due to quantum correlations~\cite{Dirac, Anderson:1973eo}.  The latter is experimentally observed in one-dimensional (1D) systems~\cite{PhysRevB.44.12361,PhysRevLett.85.832,Lake:2005tb,PhysRevLett.86.1335,*PhysRevB.68.134424,PhysRevB.67.104431,Mourigal:2013ek, PhysRevLett.84.5868,*Oosawa:1999bja,Ruegg2003}, and well described based on the Bethe ansatz~\cite{Bethe1931,PhysRevLett.111.137205}, while the former is ubiquitous in 3D systems. These two states are believed to be mutually exclusive, and hence so are their emerging spin dynamics.  On one hand, a highly entangled quantum state gives rise to non-local, multi-particle, and fractional $S=1/2$ excitations called spinons, characterized by a continuum spectrum in neutron scattering~\cite{Faddeev:1981cl, PhysRevLett.66.1529}. On the other hand, the collective excitations of ordered moments result in well-defined, single-particle $S=1$ excitations called magnons, which can be described by the linear spin-wave theory (LSWT)~\cite{hulthen,10.1103/physrev.86.694, 10.1103/physrev.87.568}.   However, recent studies reveal the shortcoming of LSWT, which was foreseen by Anderson~\cite{10.1103/physrev.86.694} and Kubo~\cite{10.1103/physrev.87.568} when they first formulated the theory seven decades ago, in describing spin dynamics emerging from classically ordered states in spin-1/2 2D edge-sharing triangular~\cite{PhysRevLett.116.087201, PhysRevLett.110.267201,Kamiya2018, Ito2017, 10.1103/physrevb.102.064421} and square lattices~\cite{Piazza:2015gi}. 

In a 2D corner-sharing-triangle kagome lattice antiferromagnet, frustration can destabilize a classical state, and a quantum spin liquid state analogous to that in the 1D systems was theoretically proposed~\cite{Balents:2010ds, Savary:2017fk,10.1126/science.aay0668} and experimentally investigated most intensely on herbertsmithite~\cite{Shores:2005de,2007PhRvL..98g7204M, PhysRevLett.98.107204, Han:2012fo,RevModPhys.88.041002}.  Even though a majority of the realizations of the kagome lattice antiferromagnet are magnetically ordered at low temperatures due to extra terms in a spin Hamiltonian, hints of quantum correlations can still be present in spin dynamics, exposing the limitation of LSWT\@. In this letter, we demonstrate the breakdown of LSWT in describing spin dynamics emerging from a classical N\'eel state in the spin-1/2 frustrated kagome lattice antiferromagnet Cs$_2$Cu$_3$SnF$_{12}$. 

\begin{figure}
\centering
\includegraphics[width=1.0\columnwidth]{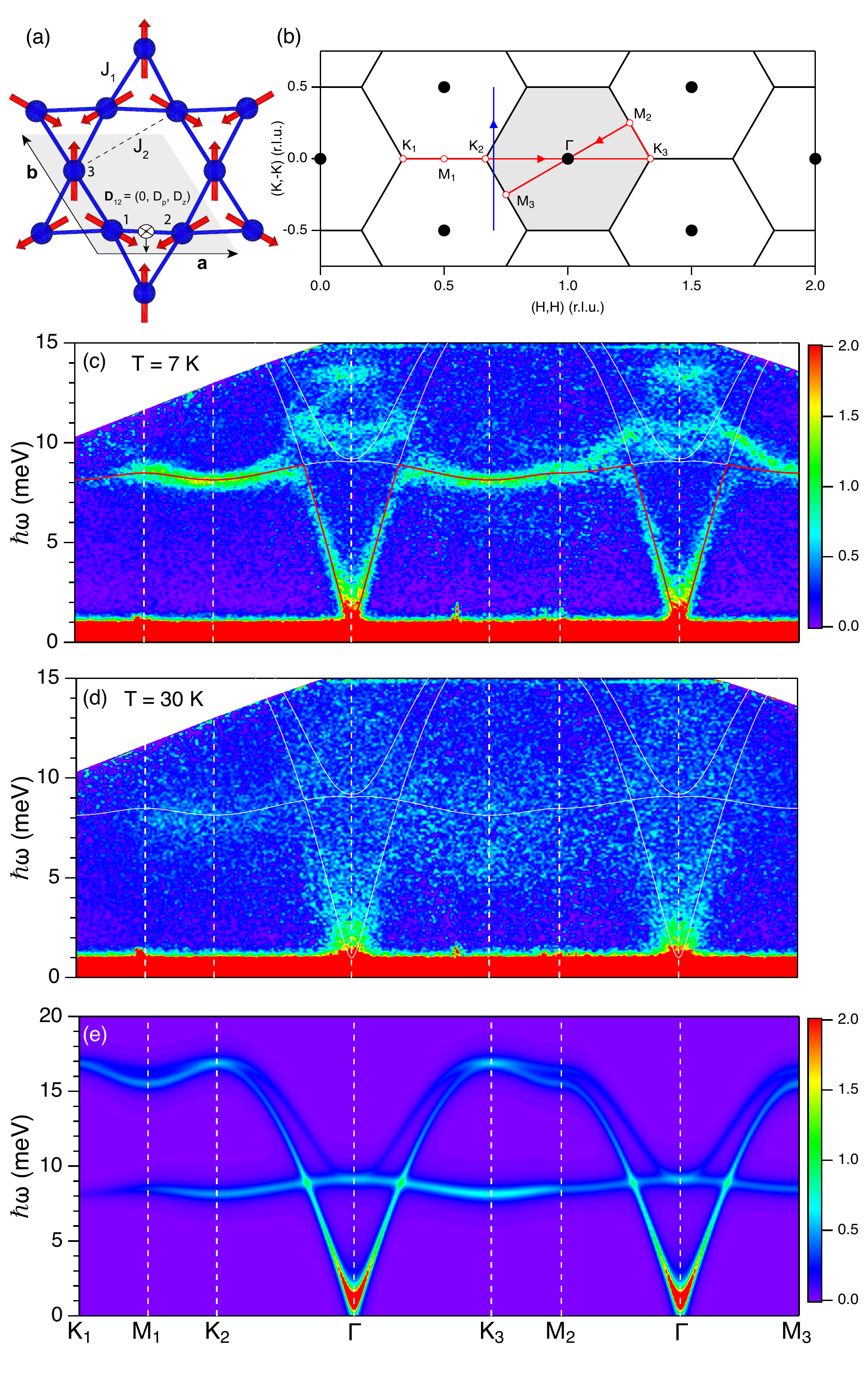}
\caption{(a) The 2D $R\bar{3}m$ unit cell (shaded area) and the all-in-all-out spin structure are depicted.  The uniform exchange interaction $J_1$ is assumed with one of the next-nearest neighbor interactions displayed. The DM vector is shown for one bond, and those for other bonds can be obtained using symmetry operators. (b) The Brillouin zones corresponding to the unit cells in (a) are depicted by the hexagons. (c)-(e) Measured scattering intensity maps as a function of the momentum transfer and energy transfer show the magnetic excitation spectrum at  (c) 7~K and (d) 30~K along the high-symmetry directions, which are depicted by the red lines in (b). For comparison, the calculated intensity map based on LSWT along the same directions is shown in (e).}\label{fig1}
\end{figure}

At room temperature, Cs$_2$Cu$_3$SnF$_{12}$ crystallizes in the rhombohedral space group $R\bar{3}m$, where the nearest-neighbor Cu$^{2+}$ ions form a network of the kagome lattice comprising corner-sharing equilateral triangles~\cite{PhysRevB.79.174407}.  A 2D unit cell contains three spins [Fig.~\ref{fig1}(a)] with its corresponding Brillouin zone depicted in Fig.~\ref{fig1}(b).  At $T_s=185$~K, the compound undergoes a structural phase transition to the monoclinic space group $P2_1/n$ causing small distortion to the triangles~\cite{PhysRevB.99.224404}.   Below $T_N=20.2$~K, the $S=1/2$ Cu$^{2+}$ spins order due to the antisymmetric Dzyaloshinskii-Moriya (DM) interaction and weak interlayer coupling, forming the all-in-all-out magnetic structure~\cite{PhysRevB.61.6156,PhysRevB.67.224435} [Fig.~\ref{fig1}(a)] with about a one-third reduction of the ordered moment~\cite{PhysRevB.99.224404}.  Previous neutron scattering measurements of magnetic excitations reveal a large negative quantum renormalization of the exchange interaction from the high-temperature value, indicative of a significant quantum effect on spin dynamics~\cite{Ono:2014kp}.

\begin{figure}
\centering \vspace{0in}
\includegraphics[width=1.0\columnwidth]{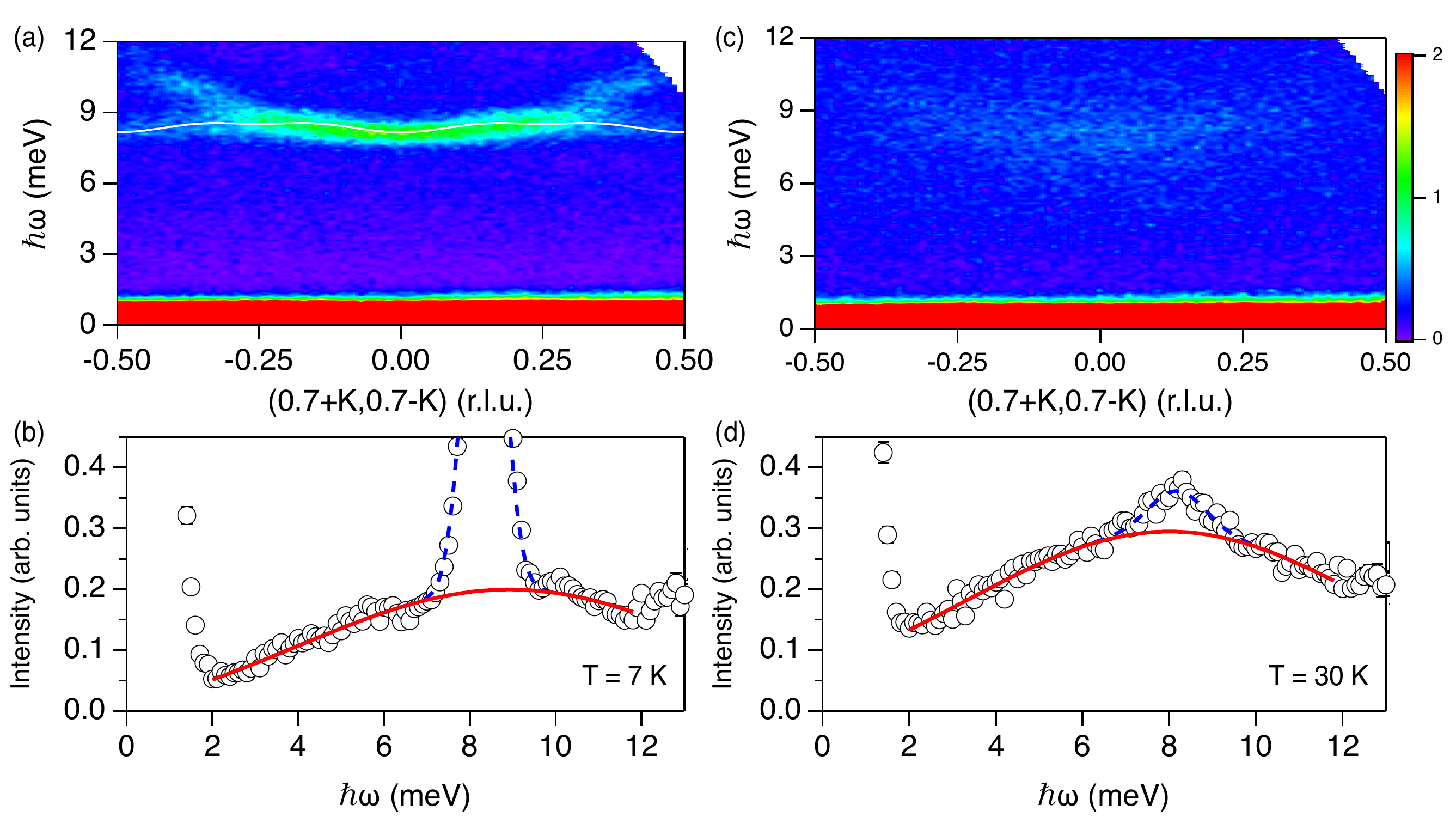}
\caption{Temperature dependence of the flat mode and broad scattering around the flat mode.  Scattering intensity maps as a function of energy and momentum transfer along $(K,-K)$ centered at $(0.7,0.7)$ as shown by the blue line in Fig.~\ref{fig1}(c) were measured at (a) 7~K and (c) 30~K. The integrated scattering intensity [$(H,H)$ = $(0.6,0.6)$ to $(0.8,0.8)$ r.l.u. and $(K,-K)$ = $(-0.25,0.25)$ to $(0.25,-0.25)$ r.l.u.] as a function of energy are shown in (b) and (d) for 7 and 30 K, respectively. The solid line in (a) shows the LSWT result. In (b) and (d), the red solid lines denote fits to a gaussian for the broad scattering and the dashed blue lines for the flat mode.}\label{fig2}
\end{figure}

High-resolution time-of-flight (TOF) inelastic neutron scattering~\cite{doi:10.1143/JPSJS.80SB.SB028} described in Supplementary Information was performed to study spin dynamics in Cs$_2$Cu$_3$SnF$_{12}$.  The measured scattering-intensity map as a function of energy transfer $(\hbar\omega)$ and momentum transfer measured at 7~K along high-symmetry directions, illustrated by the red lines in Fig.~\ref{fig1}(b), displays the energy spectrum of spin dynamics up to 15~meV.  We first attempt to describe the magnetic excitations using LSWT.  Spin-wave calculations are based on the  all-in-all-out spin structure with a uniform nearest-neighbor exchange interaction $J_1$ [Fig.~\ref{fig1}(a)].  The lattice distortion, which results in the non-uniformity of the exchange interactions, is very small~\cite{PhysRevB.79.174407, PhysRevB.99.224404}.  Hence, the distortion-induced, spatially anisotropic exchange interactions are ignored. To first approximation, the spin Hamiltonian is given by
\[
{\cal H}=\sum_{\langle i, j\rangle}\!\left[J_1\mathbf{S}_i\cdot\mathbf{S}_j+\mathbf{D}_{ij}\cdot\left(\mathbf{S}_i\times\mathbf{S}_j\right)\right] +\!\sum_{\langle l, k\rangle}\!J_2\mathbf{S}_l\cdot\mathbf{S}_k,
\]
where $J_2$ is the next-nearest-neighbor interaction and $\mathbf{D}_{ij}$ denotes the DM vector between the nearest-neighbor spins.  $\mathbf{D}_{12}=(0,D_p,D_z)$ is shown in Fig.~\ref{fig1}(a) and all other DM vectors can be obtained by symmetry operators. 

\begin{figure*}
\centering \vspace{0in}
\includegraphics[width=0.825\paperwidth]{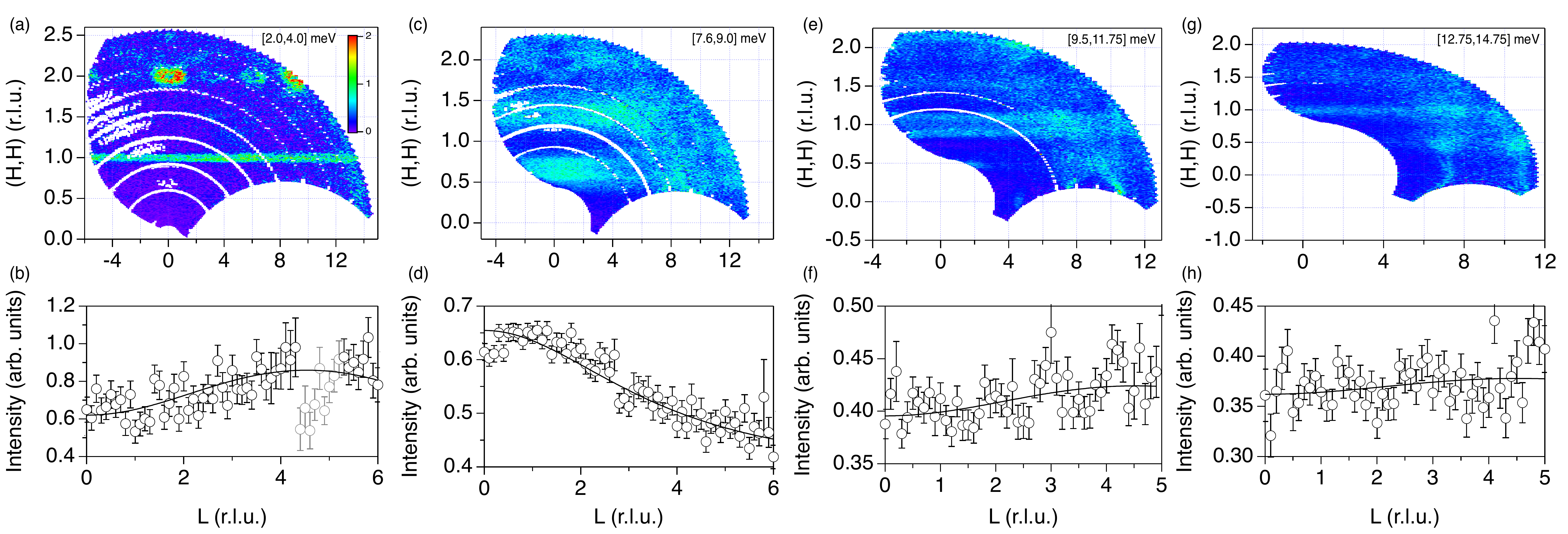}
\caption{Scattering intensity maps as a function of $(H,H)$ and $L$ measured at 7~K show rod-like scattering intensity along $L$ for the integrated energy ranges (a) $[2.0, 4.0]$~meV, (c) $[7.6, 9.0]$~meV, (e) $[9.5, 11.75]$~meV, and (g) $[12.75, 14.75]$~meV.  (b), (d), (f), and (h) show integrated intensity as a function of $L$.  The integrated ranges for $(H,H)$ are (b) $(0.95, 0.95)$ to $(1.05,1.05)$, (d) $(0.450, 0.450)$ to $(0.825,0.825)$, (f) $(0.8, 0.8)$ to $(1.2,1.2)$, and (h) $(0.9, 0.9)$ to $(1.1,1.1)$ r.l.u.  Solid lines in (b), (d), (f), and (g) denote fits to the polarization factor described in the text.}\label{fig3}
\end{figure*}

\begin{figure}
\centering \vspace{0in}
\includegraphics[width=1.0\columnwidth]{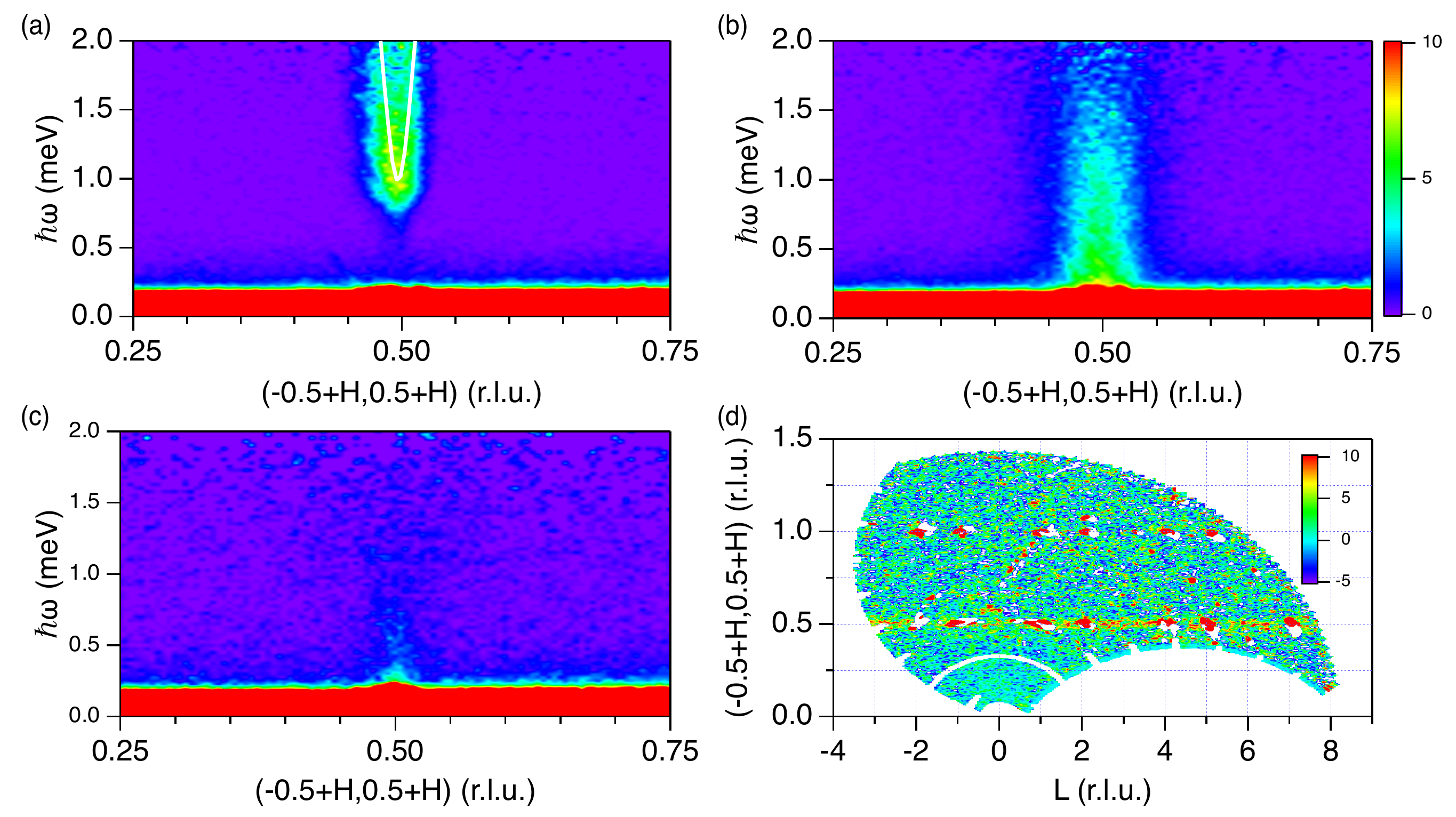}
\caption{Temperature dependence of the spin-anisotropic energy gap.  Scattering intensity maps as a function of energy and momentum transfer along $(H, H)$ centered at (0,1) show (a) the energy gap at 7 K, (b) the closing of the gap at 30 K with the remnant scattering intensity at low energy, and (c) the disappearance of the remnant scattering at 150 K.  The solid line in (a) denotes the LSWT result.  In (d), the difference map between the integrated quasi-elastic intensity $I_{\textrm{q-el}}$ (an integrated energy range of $[-0.5,0.5]$~meV) measured at 30 K and 7 K $[I_{\textrm{q-el}}(30K)-I_{\textrm{q-el}}(7K)]$ shows the scattering rod at (0,1) along $L$.} \label{fig4}
\end{figure}

The spin-wave calculations exhibit three transverse magnon modes, two of which, the `weathervane' (flat)~\cite{1992PhRvB..4512377S,1992PhRvB..45.2899H,10.1103/physrevb.45.7536}  and dispersive low-energy modes, were fitted to the measured dispersion to obtain $D_z$ and $J_1$, respectively.  The weakly dispersive nature of the flat mode is a result of ferromagnetic $J_2$.  $D_p$, which determines the spin-wave anisotropic gap at the zone center [see also Fig.~\ref{fig4}(a)], was fit to reproduce this gap energy.  The best fit denoted by the red line in Fig.~\ref{fig1}(c) yields the fit parameters $J_1=13.3(2)$, $J_2=-0.24(6)$, $D_p=0.566(2)$, and $D_z=-1.94(8)$~meV.  The value of $J_1$ represents the renormalization factor of 0.67, consistent with the previous work~\cite{Ono:2014kp}. The spin-wave results [see Figs.~\ref{fig1}(c) and \ref{fig1}(e)] can capture the low-energy part of the resolution-limited, single-particle magnetic excitations up to the flat mode around 8 meV.   However, it fails to reproduce the full spectrum.  Comparing the measured dispersion in Fig.\ref{fig1}(c) and the calculated one in Fig.~\ref{fig1}(e), we highlight the inconsistencies in (i) the absence of the flat mode around the zone center, (ii) the absence of the high-energy dispersive modes around the $M$ and $K$ points [see also Figs.~2(a) and 2(b) in Ref.~\onlinecite{Saito}], (iii) three weakly dispersive, resolution-limited modes centered at 9.6(2), 10.7(1), and 13.4(1)~meV measured at the $\Gamma$ point, and (iv) broad scattering around the flat mode [see also Fig.~\ref{fig2}(a)]. Furthermore, we note that even though LSWT appears to well reproduce the lower branch of the flat mode along the zone edges but it fails to capture the branching out to high energy near the zone center as observed in Fig.~\ref{fig1}(c) [see also Fig.~\ref{fig2}(a)], which suggests the existence of another mode beyond LSWT.  

The most conspicuous discrepancy between the measured dispersion and LSWT is the existence of the three weakly dispersive modes around the zone center between 9 and 14 meV.  The spin-wave calculations predict only one highly dispersive magnon mode starting from 9.2 meV at the $\Gamma$ point and reaching 16.9 meV at the $K$ point [Fig.~\ref{fig1}(e)]. Unlike the flat and dispersive low-energy modes, which have some parts that agree with the measured dispersion, this magnon mode is entirely absent.  Instead, it is replaced by three weakly dispersive modes, which quickly terminate away from the zone center.  The termination wave vector approaches the zone center for the higher-energy modes forming a top-of-pyramid-like profile [Fig.~\ref{fig1}(c)].  We note that these modes were not observed in the spin-5/2 kagome lattice antiferromagnet KFe$_3$(OH)$_6$(SO$_4$)$_2$ (jarosite)~\cite{PhysRevLett.96.247201} despite both having the same classical all-in-all-out ordered state, suggesting the manifestation of the quantum effect on spin dynamics in Cs$_2$Cu$_3$SnF$_{12}$. 

Another distinct feature of the measured dispersion that cannot be reproduced by LSWT is the broad scattering around the flat mode, which is most evident in the intensity map along the $[K, -K]$ direction centered at $(0.7,0.7)$ as shown in Figs.~\ref{fig2}(a) and ~\ref{fig2}(c) for 7~K and 30~K, respectively. This broad intensity is highlighted in the integrated intensity as a function of energy by the red solid lines in Figs.~\ref{fig2}(b) and \ref{fig2}(d).  The onset of the broad intensity, which extends up to 12~meV, starts around 2~meV.  At 7~K (30~K), it peaks at 8.9(2) [8.0(1)]~meV, nearly coinciding with the flat mode centered at 8.32(1) [8.2(1)]~meV.  The proximity of the single-particle flat mode to the broad scattering could hint at their interconnection.

Analysis of the spin-fluctuation polarization was performed for the low-energy dispersive mode, flat mode, and three modes between 9 and 14 meV.  The polarization factor in the magnetic scattering cross-section is governed by $f(Q)^2\left[1\pm(Q_L/Q)^2\right]$, where $f(Q)$ is the Cu$^{2+}$ magnetic form factor, the positive (negative) sign is for in-plane (out-of-plane) polarization, and $Q_L$ ($Q$) denotes the $L$ component (magnitude) of ${\bf Q}$.  Figures~\ref{fig3}(a), \ref{fig3}(c), \ref{fig3}(e), and \ref{fig3}(g) show the constant-energy contour maps in the $(H,H,L)$ plane, where magnetic scattering form a rod along $L$ attesting to the two-dimensionality of the spin network. The integrated intensity along the scattering rod as a function of $L$ for the low-energy dispersive mode [Fig.~\ref{fig3}(b)] reveals the in-plane polarization, consistent with the fact that the anisotropic gap at the bottom of this mode [Fig.~\ref{fig4}(a)] results from the in-plane component $D_p$ of the DM vector, which functions as effective easy-axis anisotropy.  On the other hand, as shown in Fig.~\ref{fig3}(d) the flat mode is out-of-plane polarized, confirming its connection to the out-of-plane component $D_z$, which functions as effective easy-plane anisotropy.  The polarization of the three modes around the zone center [Fig.~\ref{fig3}(f) for the two modes at 9.6 and 10.7~meV and Fig.~\ref{fig3}(h) for the 13.41-meV mode] suggest that they are in-plane polarized, which could indicate that these modes are longitudinal modes given that quasi-particles are confined within the kagome plane.  The polarization analysis appears to suggest that the in-plane modes only exist around the zone center while the out-of-plane mode disappears near the zone center but prevails near the zone edges. Furthermore, while the calculated spin-wave results indicate that the out-of-plane-polarized flat mode attains high intensity throughout the Brillouin zone and the in-plane-polarized dispersive mode survive up to high energies [Fig.~\ref{fig1}(e)], the data [Fig.~\ref{fig1}(c)] show that both modes terminate where they cross each other. 

At 30~K, about 10~K above $T_N$, the resolution-limited, single-particle excitations of the low-energy dispersive mode and flat mode, which is dominant near the zone edges, disappear and are replaced by a column of continuum scattering [Fig.~\ref{fig1}(d) and \ref{fig4}(b)] and a broad peak [indicated by the dashed blue line in Fig.~\ref{fig2}(d)], respectively.  The remnant broad scattering of the flat mode, which remains centered around 8~meV, implies that spins are confined to fluctuate within the kagome plane and retain two-dimensional rotational symmetry in the plane, while the closing of the in-plane anisotropy gap [Fig.~\ref{fig4}(b)] suggests that spins can freely rotate in the plane.  Furthermore, due to the effect of the out-of-plane DM interaction, which determines spin chirality, the in-plane fluctuations, which form a scattering rod along $L$ as shown in Fig.~\ref{fig4}(d) and disappear at 150 K [Fig.~\ref{fig4}(c)], must preserve the chirality of the ordered state.  Therefore, above $T_N$, the system transitions into a chiral ordered state with the absence of rotational symmetry breaking. This chiral state was previously observed in jarosite~\cite{Grohol:2005bg}.  The low-energy spin fluctuations resulting in the quasi-elastic scattering observed in Figs.~\ref{fig4}(b) and \ref{fig4}(d) can be associated with a critical phenomenon of the ordered moments at the magnetic phase transition, supporting the dominance of magnons at low energy, which could also explain why LSWT works exceptionally well in this regime.

However, LSWT alone cannot fully describe spin dynamics in Cs$_2$Cu$_3$SnF$_{12}$.  We can also rule out the effect of the doubling of the magnetic unit cell due to the structural transition at $T_s$ as the `ghost mode'~\cite{PhysRevB.89.024414} due to the zone folding cannot describe the discrepancies (see Supplementary Information).  Another explanation, which includes higher-order (cubic and quartic) terms in the spin-wave expansion, was examined in Ref.~\onlinecite{PhysRevB.92.094409, *PhysRevB.92.144415}, where the absence or broadening of the dispersive excitations above the flat mode in the kagome antiferromagnet can be explained by a decay of the quasi-particle into two magnons belonging to the flat mode, which is possible for a non-collinear system.  However, this scenario also fails to capture the loss of intensity of the dispersive and flat modes after their crossing as observed in Fig.~\ref{fig1}(c).  Given that the energy of the flat mode is around 8~meV, one would expect that the dispersive mode becomes broadened around 16~meV, which is twice the flat-mode energy.  However, from the data, we start to observe the loss of intensity of the dispersive mode above 9~meV.  

Furthermore, this model is unable to explain the three weakly dispersive modes around the zone center.  These modes do not resemble any of the calculated spin-wave modes, and hence their origin is most likely not due to the minor improvement on the spin-wave theory such as the inclusion of higher-order terms in the spin-wave expansion.  The fact that the peak profile of these excitations is resolution-limited (see Fig. S6 in Supplementary Information) suggests that they do not result from the broadening of the magnon modes.  We believe that these modes correspond to two-spinon bound states (triplons)~\cite{Kohno:2007ik, Nawa:2020hm}, which could decay into free spinons away from the zone center.  At 30~K, these modes also decay, forming continuum scattering extending up to 15~meV (the highest measuring energy) [Fig.~\ref{fig1}(d) and Fig.~S5 in Supplementary Information].  The spinon continuum at 30~K retains the top-of-pyramid profile, suggesting the energy conservation of the decay.  Furthermore, the upper branch of the flat mode, which disperses to high energy close to the zone center, could also be attributed to the triplons, and the decay of these triplons at 30~K could increase the continuum intensity around the flat mode [comparing the red lines in Figs.~\ref{fig2}(b) and \ref{fig2}(d)].  The fact that the center of the continuum at 30~K is roughly the same as the energy of the flat mode suggests the energy-conservative decay of the triplons into free spinons.  In addition, it was recently reported that the continuum scattering also exists up to 50~meV~\cite{Saito}, substantiating the existence of free spinons at high energy.  Therefore, we conjecture that the triplons are dominant around the zone center between 9 and 14~meV, and that they become unstable and decay into free spinons away from the zone center, at high energy, or high temperature.  It is not clear what interaction can bind pairs of spinons to form triplons.  In a spatially anisotropic triangular antiferromagnet, it was shown that a pair of spinons can lower their kinetic energy by forming a triplon that propagates between chains~\cite{Kohno:2007ik}.  

In Cs$_2$Cu$_3$SnF$_{12}$, the existence of spinons and their bound states could be attributed to the one-third of magnetic moments that remain disordered at low temperature and possibly form an entangled quantum state, whereas the dominance of magnons at low energy could be the contributions from the two-third of magnetic moments that order. In the triangular and square lattice spin-1/2 antiferromagnets~\cite{PhysRevLett.116.087201, Kamiya2018, Ito2017, Piazza:2015gi}, the deviation of the observed magnetic excitation spectrum from a result of LSWT was also observed. To explain the excitation spectra in these systems, a fermionic resonating valence-bond theory was proposed and single-particle spin dynamics arises as the two-spinon bound state~\cite{Piazza:2015gi, PhysRevX.9.031026, PhysRevB.102.075108}.  A similar mechanism could also play a key role in spin dynamics of the spin-1/2 kagome antiferromagnet.

We have performed high-resolution TOF inelastic neutron scattering and mapped out the magnetic excitations in the kagome lattice antiferromagnet Cs$_2$Cu$_3$SnF$_{12}$.  Low-energy spin dynamics can be well described by LSWT\@.   However, the theory becomes inadequate to describe the whole energy spectrum.  We observed the disappearance of the flat mode intensity near the zone center, the absence of the dispersive modes around the $M$ and $K$ points, the emergence of extra modes between 9 and 14~meV, and the broad continuum around the flat mode, all of which cannot be accounted for by LSWT\@.  Our results reveal a shortfall of the semi-classical framework, which has been ubiquitously used to describe spin dynamics of a magnetically ordered system, in describing magnetic excitations in this 2D frustrated kagome system.  The breakdown of LSWT necessitates a new quantum-based theoretical framework to describe spin dynamics in Cs$_2$Cu$_3$SnF$_{12}$ and other magnetically ordered 2D systems. 

\begin{acknowledgments}
We would like to thank H.~Tanaka for fruitful discussion and for providing us with high-quality samples. Work at Mahidol University was supported in part by the National Research Council of Thailand Grant Number N41A640158 and the Thailand Center of Excellence in Physics (ThEP)\@. The experiment on AMATERAS was performed with the approval of J-PARC (Proposal No. 2012A0090).
\end{acknowledgments}

%

\clearpage
\pagebreak
\widetext
\begin{center}
\textbf{\large Supplemental Information:\\Breakdown of Linear Spin-wave Theory and Existence of Spinon Bound States in the Frustrated Kagome Lattice Antiferromagnet}
\end{center}
\setcounter{equation}{0}
\setcounter{figure}{0}
\setcounter{table}{0}
\setcounter{page}{1}
\makeatletter
\renewcommand{\theequation}{S\arabic{equation}}
\renewcommand{\thefigure}{S\arabic{figure}}

\begin{figure}[htp]
\centering \vspace{0in}
\includegraphics[width=0.4\textwidth]{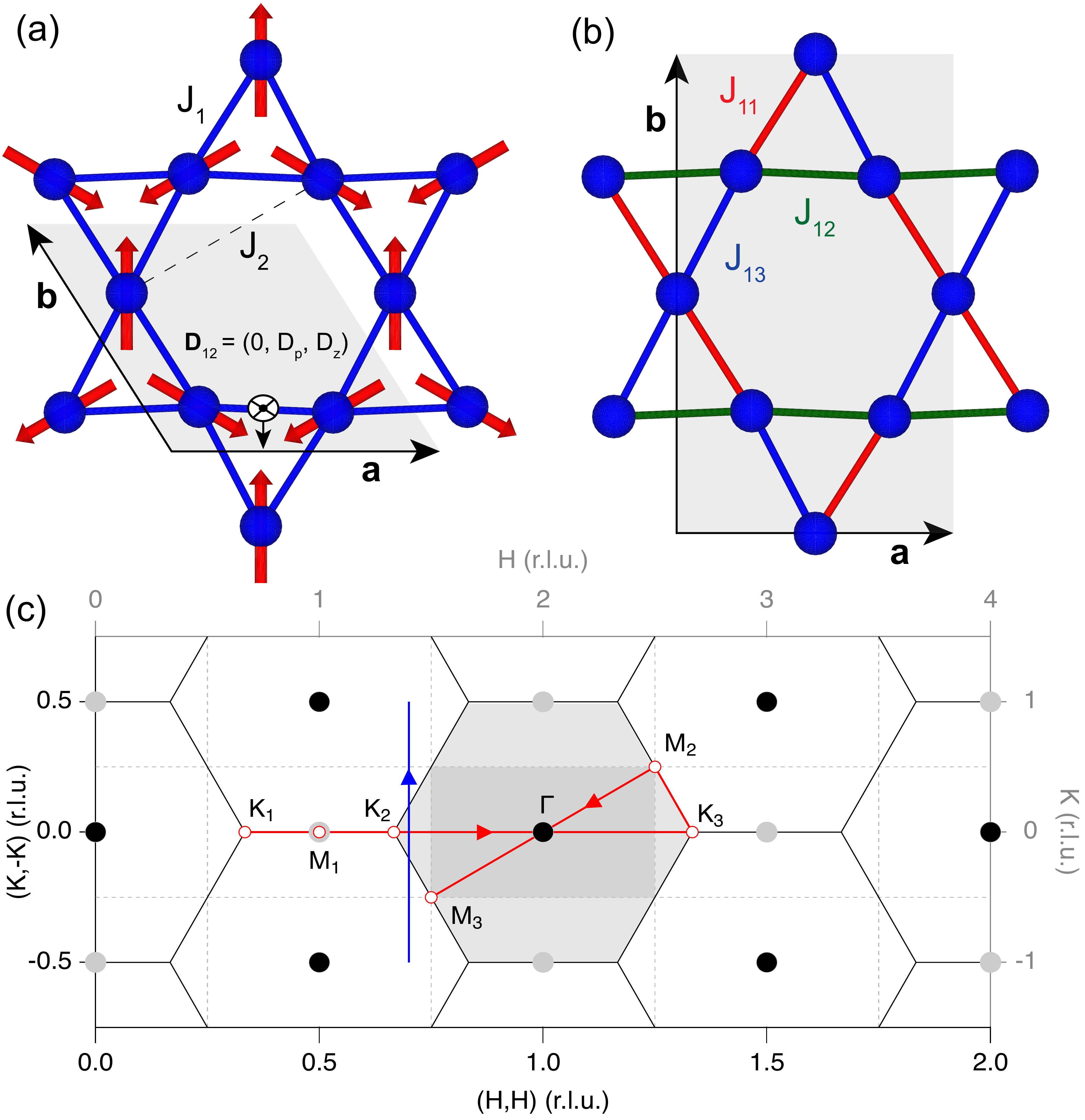}
\caption{(a) The 2D $R\bar{3}m$ unit cell (shaded area) and the all-in-all-out spin structure are depicted.  The uniform exchange interaction $J_1$ is assumed with one of the next-nearest neighbor interaction displayed. The DM vector is shown for one bond, and those for other bonds can be obtained using symmetry operators. (b) The 2D unit cell projected on to the kagome plane for $P2_1/n$ is shown by the shaded area.  Three inequivalent exchange interactions are denoted by $J_{11}$, $J_{12}$, and $J_{13}$.  (c) The Brillouin zones corresponding to the unit cells in (a) [(b)] are depicted by the hexagonal (rectangular) areas.}\label{figS1-1}
\end{figure}

\noindent
{\it \bf Inelastic neutron scattering measurements}\\[1mm]
Single-crystal  Cs$_2$Cu$_3$SnF$_{12}$ was synthesized using a method described in Ref.~\cite{PhysRevB.79.174407}.  Magnetic excitations in Cs$_2$Cu$_3$SnF$_{12}$ were measured on the cold-neutron time-of-flight disk-chopper spectrometer AMATERAS~\cite{doi:10.1143/JPSJS.80SB.SB028} at Japan Proton Accelerator Research Complex.  The sample, which is consisted of three co-aligned single crystals of the total mass 3.82~g, was aligned so that the $[110]$ and $[001]$ directions are horizontal and in the scattering plane. Reciprocal lattice vectors throughout this article are given in the rhombohedral space group $R\bar{3}m$ of the high-temperature phase, where the 2D magnetic unit cell is denoted by the shaded area in Fig.~\ref{figS1-1}(a), with the lattice parameter $a=7.105$ and $c=20.381$~\AA, and the corresponding Brillouin zone by the solid lines in Fig.~\ref{figS1-1}(c). The sample was loaded into an aluminum can and cooled down to a base temperature of 7 K using a closed cycle $^4$He cryostat. The monochromator disk choppers rotated at 150 Hz with 30 mm width slits and two different conditions for choppers (including auxiliary choppers) to select two different set-ups of incident energies of (i) 4.0, 6.4, 11.7, and 27.6~meV and (ii)  5.0, 8.4, 17.0, and 51.0~meV. We found that the incident energy of 17.0~meV gives the optimal results in terms of dynamic-range coverage and incident neutron flux. The resolution of the incident energy of 5.0 meV is sufficient to resolve the magnetic-anisotropy gap of ~0.7 meV at the magnetic zone center.  Multiple datasets were acquired by rotating the sample about the vertical axis, which is parallel to the $(\bar{1}10)$ direction, in a step of $2^\circ$ covering roughly 100$^\circ$ of the sample orientation.  The magnetic excitations were measured at the base temperature and 30 K, which is about 10~K above $T_N$, while phonon background was measured at 150 K.  These data were processed using the software package \texttt{Utsusemi}~\cite{doi:10.7566/JPSJS.82SA.SA031} to generate the four-dimensional scattering-intensity data $I(\mathbf{Q},\hbar\omega)$, where $\mathbf{Q}$ is the momentum transfer and $\hbar\omega$ is the energy transfer.  The obtained data were then sliced and cut along high symmetry directions to produce scattering intensity maps and line scans.  Taking advantage of the non-dispersive and rod-like scattering along the direction perpendicular to the kagome plane, which results from the two-dimensionality of the system, we integrate the intensity along the $[001]$ direction to increase a signal-to-background ratio.  To avoid phonon contributions, the $L$ integration for most of the figures in the main text was taken for $L$-range of $[-3, 3]$~r.l.u. except for the $H-K$ maps, where the integration was taken for the whole measuring $L$ range.\\[3mm]

\begin{figure}[htp]
\centering \vspace{0in}
\includegraphics[width=0.6\textwidth]{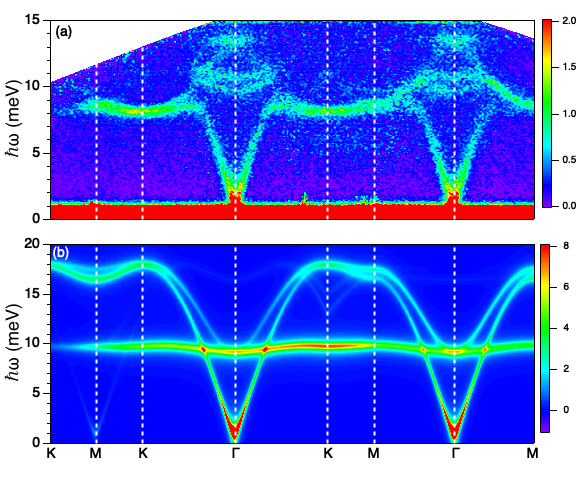}
\caption{Intensity maps as a function of momentum and energy show a comparison between the observed excitation spectrum and calculated spin-wave spectrum based on the enlarged 2D magnetic unit cell. We note that the calculations were done for one domain. Therefore, the ``ghost mode'' only appears at one $M$ point on the left.  However, the structural phase transition gives rise to a total of three domains, and hence the ``ghost-mode'' excitations appear at all $M$ points.}\label{figS1-2}
\end{figure}

\noindent
{\bf Enlarged magnetic unit cell and folding of excitation spectrum}\\[2mm]
The enlarged magnetic unit cell resulting from the structural phase transition from $R\bar{3}m$ to $P2_1/n$  shown in Fig.~\ref{figS1-1} cannot explain the magnetic excitation spectrum above the flat mode.  Figure~\ref{figS1-2} shows a comparison between the data and calculated spin-wave spectrum based on the enlarged 2D magnetic cell containing 6 spins.  This enlarged magnetic cell is the projection of a monoclinic unit cell in space group $P2_1/n$ onto the kagome plane.  The enlarged unit cell gives rise to the folding of the Brillouin zone as evidenced by an extra dispersive mode originating from one of the M-points.  This mode is very weak due to the all-in-all-out magnetic structure, which yields weak magnetic scattering intensity at the M-point.  In particular, the folding does not give rise to the three modes between 9 and 14~meV around the $\Gamma$ point. \\[3mm]

\begin{figure}[htp]
\centering \vspace{0in}
\includegraphics[width=0.8\textwidth]{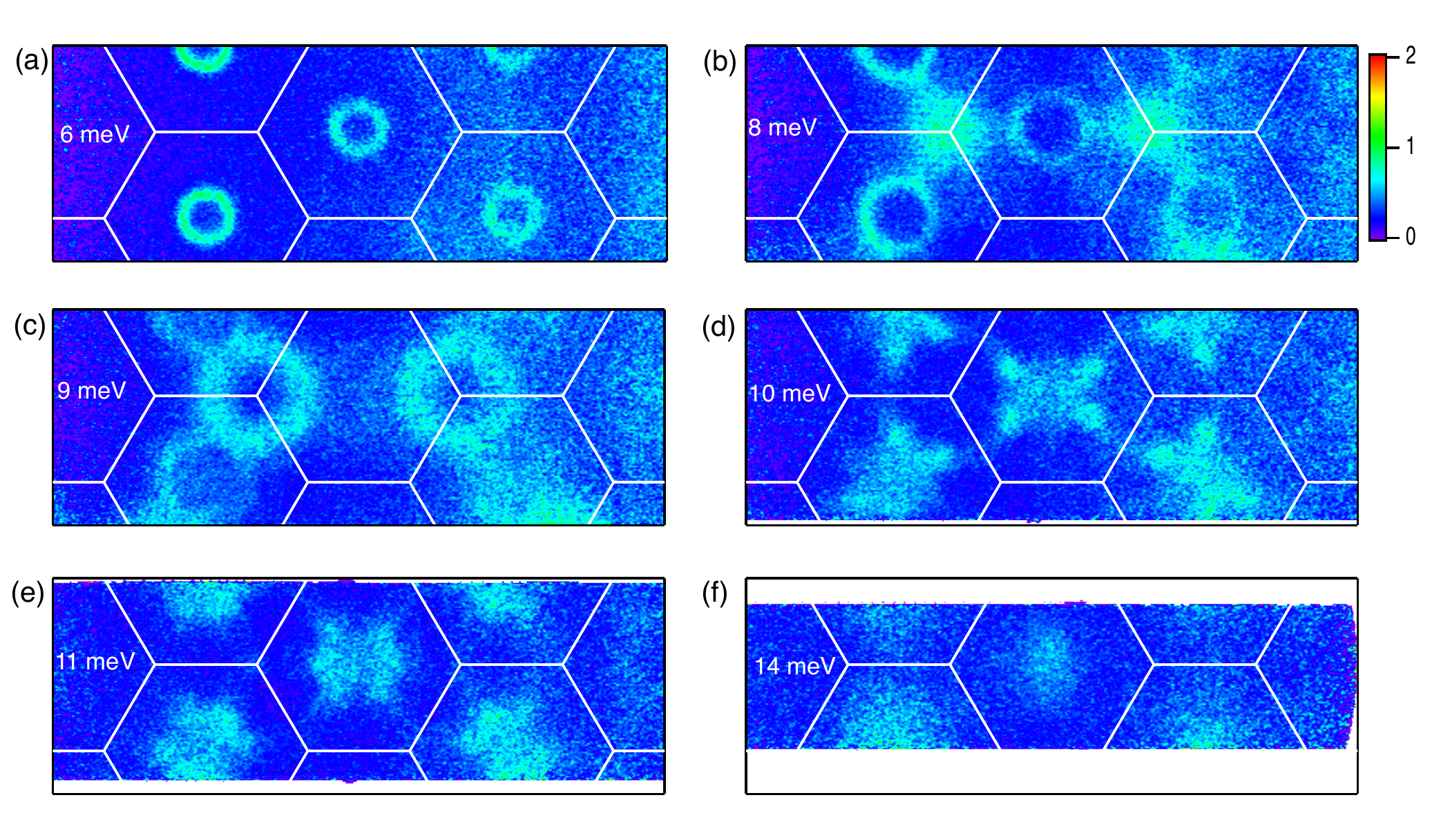}
\caption{Measured scattering intensity maps at a constant energy as a function of in-plane momenta ($H-K$ plane) for (a) 6, (b) 8, (c) 9, (d) 10, (e) 11, and (f) 14~meV.}\label{figS2-1}
\end{figure}

\begin{figure}[htp]
\centering \vspace{0in}
\includegraphics[width=0.8\textwidth]{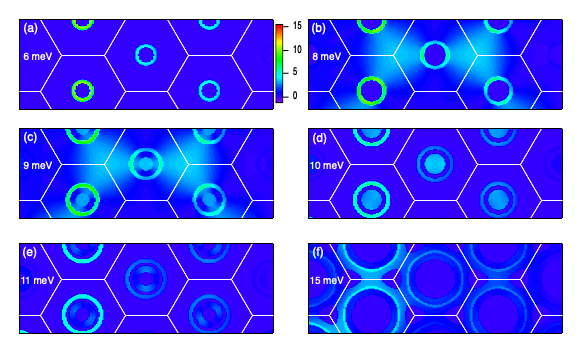}
\caption{Scattering intensity maps at a constant energy as a function of in-plane momenta ($H-K$ plane) were calculated based on the linear spin-wave theory for (a) 6, (b) 8, (c) 9, (d) 10, (e) 11, and (f) 15~meV.}\label{figS2-2}
\end{figure}

\begin{figure}[!htp]
\centering \vspace{0in}
\includegraphics[width=0.9\textwidth]{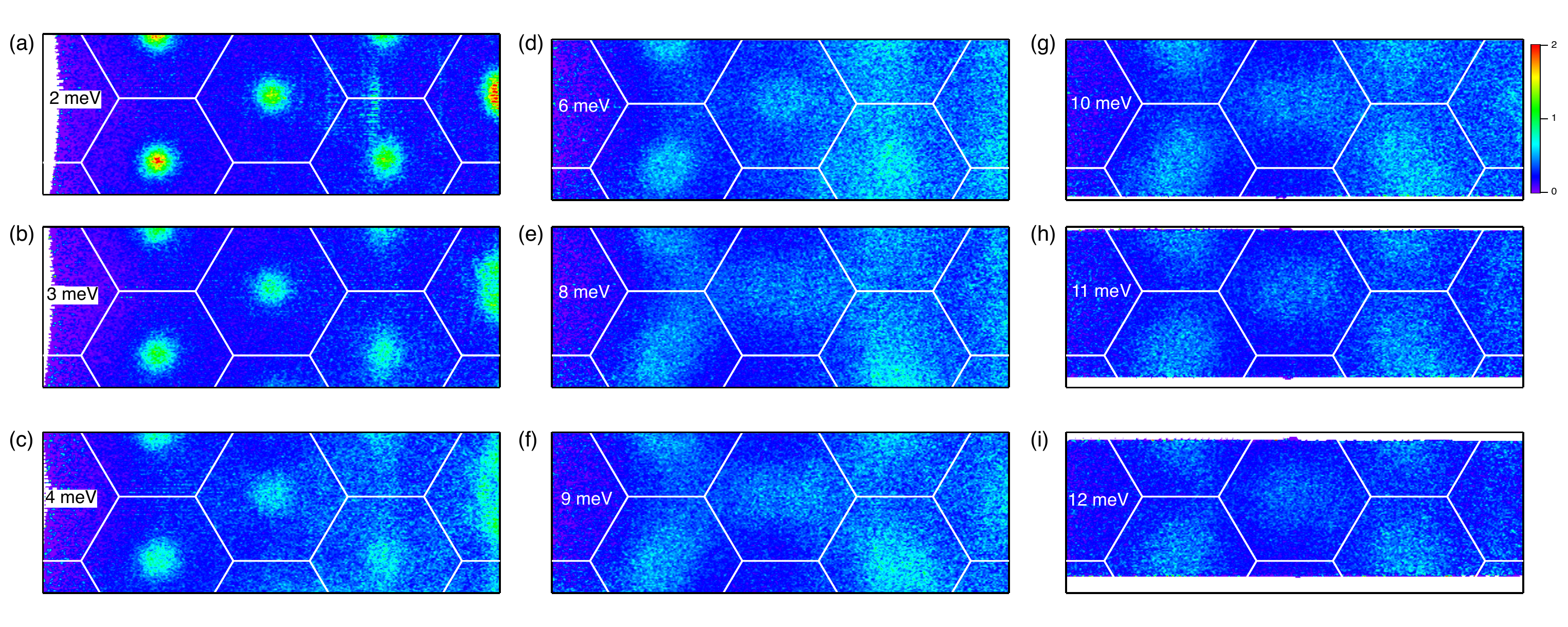}
\caption{Constant-energy maps show the magnetic excitations in the hk0 plane measured at 30~K. The intensity was integrated along the [001] direction. The data shown in (a)-(i) were integrated over the energy range of [$\hbar\omega-0.5, \hbar\omega+0.5$] for $\hbar\omega$ of 2, 3, 4, 6, 8, 9, 10, 11, and 12 meV, respectively.}\label{figS5}
\end{figure}

\noindent
{\bf Intensity maps in $H-K$ plane}\\[1mm]
Comparing the measured intensity maps in the $H-K$ plane shown in Fig.~\ref{figS2-1} with the calculated ones based on the linear spin-wave theory shown in Fig.~\ref{figS2-2}, we found that the observed intensity maps agree well with the LSWT results at 6 and 8~meV.  The discrepancy starts to appear at 9~meV, where the measured data show that the scattering intensity forms a ring around the point where corners of the three zones meet but the calculated results show that the majority of the intensity remains around the zone center.  At 10 and 11 meV, while the calculations show the scattering intensity centered around the zone center forming a circular pattern, the data around the zone center show cross and star-shape patterns.  Furthermore, the calculations suggest that the spin-wave intensity only extends up to $~\sim18$~meV, where the majority of the scattering intensity is near the zone boundary, which is inconsistent with the data.  The measured intensity maps measured at 30 K are also shown in Fig.~\ref{figS5}.\\[3mm]

\clearpage

\begin{figure}[!htp]
\centering \vspace{0in}
\includegraphics[width=0.9\textwidth]{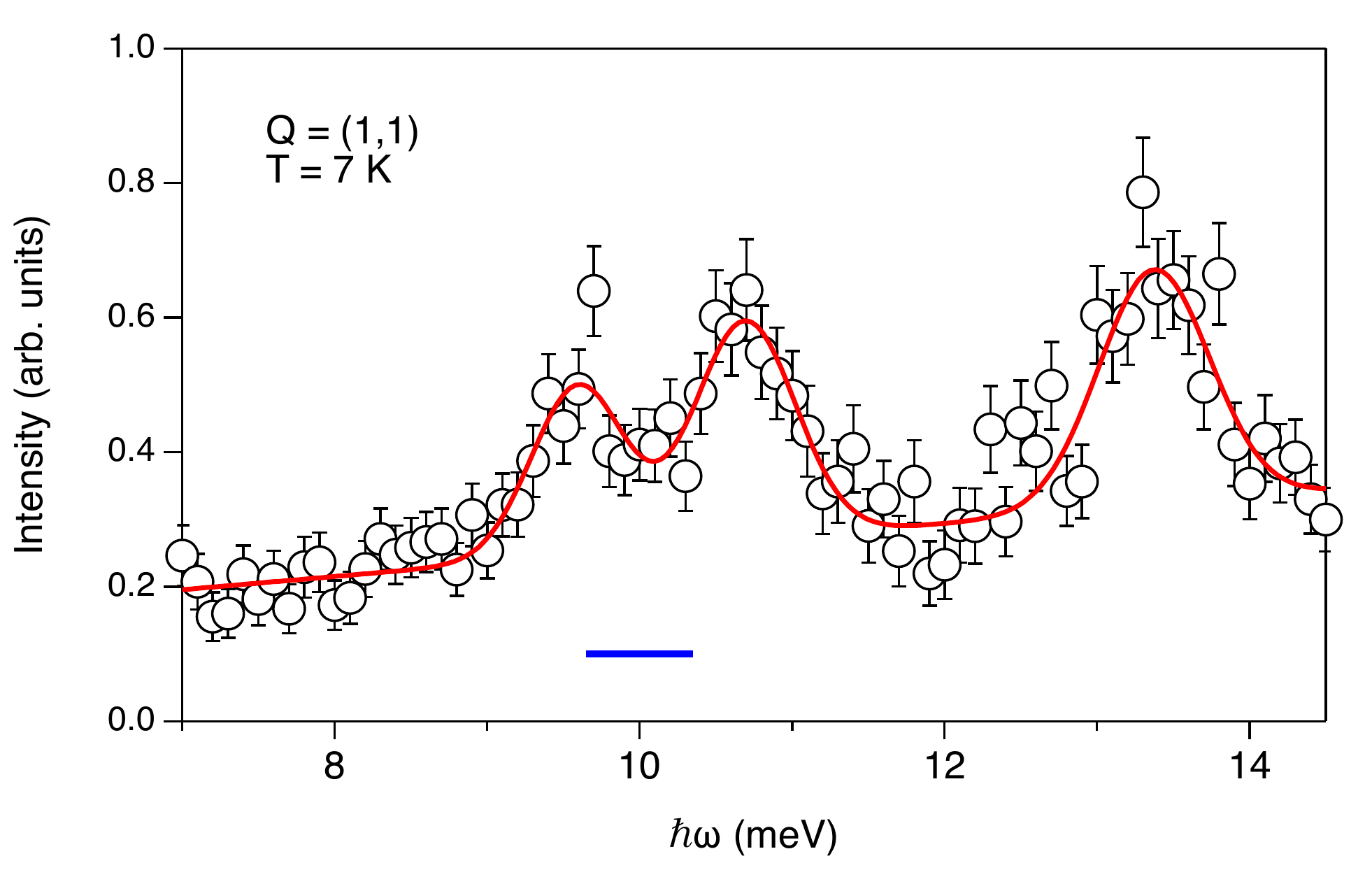}
\caption{An energy-cut through the zone center at 7~K shows the three weakly dispersive modes centered at 9.6(2), 10.7(1), and 13.4(1)~meV. The horizontal blue line denotes the energy resolution measured at the elastic position.}\label{figS6}
\end{figure}

\end{document}